\documentclass[aps,prl,twocolumn,superscriptaddress,showpacs]{revtex4}
\usepackage{graphicx}
\usepackage{times}
\begin{document}
\def\T{\Theta}
\def\D{\Delta}
\def\d{\delta}
\def\r{\rho}
\def\p{\pi}
\def\a{\alpha}
\def\g{\gamma}
\def\ra{\rightarrow}
\def\s{\sigma}
\def\b{\beta}
\def\e{\epsilon}
\def\G{\Gamma}
\def\om{\omega}
\def\pe{$1/r^\a$ }
\def\l{\lambda}
\def\f{\phi}
\def\w{\psi}
\def\m{\mu}
\def\t{\tau}
\def\dg{d^3{\bf r}\,d^3{\bf p}}
\def\df{f({\bf r, p})}
\def\dn{n({\bf r, p})}

\title{Comment on ''Microcanonical Mean Field Thermodynamics of 
Self-Gravitating and Rotating systems''}
\author{I.~Ispolatov}
\affiliation{Departamento de Fisica, Universidad de Santiago de Chile,
Casilla 302, Correo 2, Santiago, Chile}
\author{M.~Karttunen}
\affiliation{Biophysics and Statistical Mechanics Group, 
Laboratory of Computational Engineering, Helsinki University
of Technology, P.O. Box 9203, FIN-02015 HUT, Finland}
\date{\today}
\maketitle
In a recent Letter \cite{vot} Votyakov {\it et al.} stated that a
microcanonical system of
self-gravitating particles confined to a spherical container collapses at
a sufficiently low energy and high angular momentum not to a single-core
core-halo system but to a double-core binary. Based on molecular 
dynamics simulations and analytical estimates we found strong evidence that 
such systems do not form stable or even long-living metastable binaries. 
Instead, they form  single core axially 
asymmetric states with the core traveling
on an almost circular orbit along the container wall.

We considered systems consisting of $N=$250--1000 particles of unit mass 
confined to a spherical container of radius $R$ with reflecting walls and 
interacting via soft Coulomb potential $-(r^2+r_0^2)^{-1/2}$, $r_0=0.002R$.
Details of the MD simulation setup are described in \cite{us}. 
Two types of initial configurations were used:
A spherically symmetric core-halo state and an axially
symmetric ring.
We chose the values of rescaled energy $\e=E R/ N^2$
and angular momentum $\lambda=L/\sqrt{(RN^3)}$ in the intervals 
$-1\leq \e \leq -0.4$ and $0.6\leq \lambda \leq 1.2$, where according
to Ref.~\cite{vot} double-core structures exist.
\begin{figure}
\includegraphics[width=.2\textwidth, height=.2\textwidth]{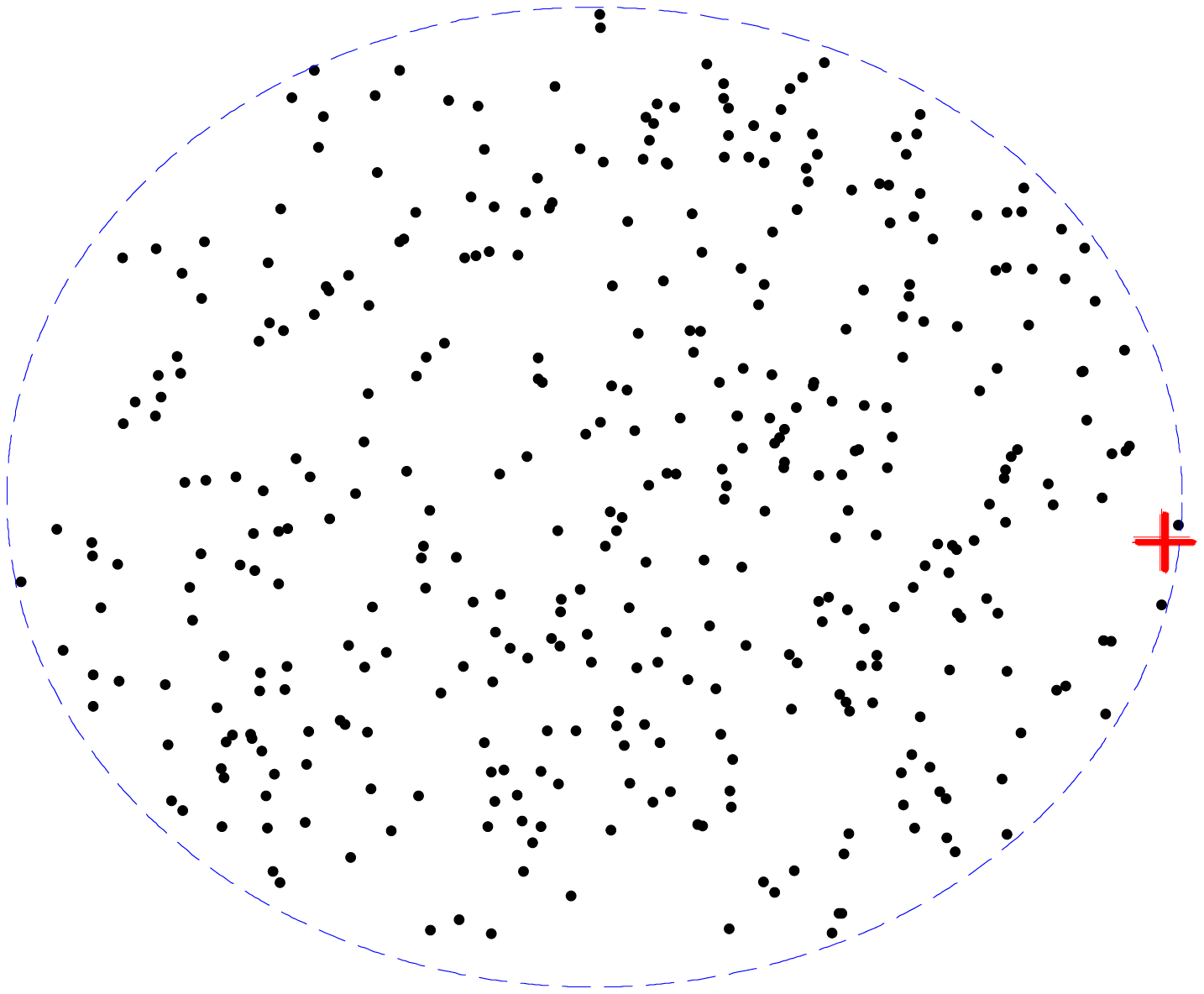}
\includegraphics[width=.2\textwidth, height=.2\textwidth]{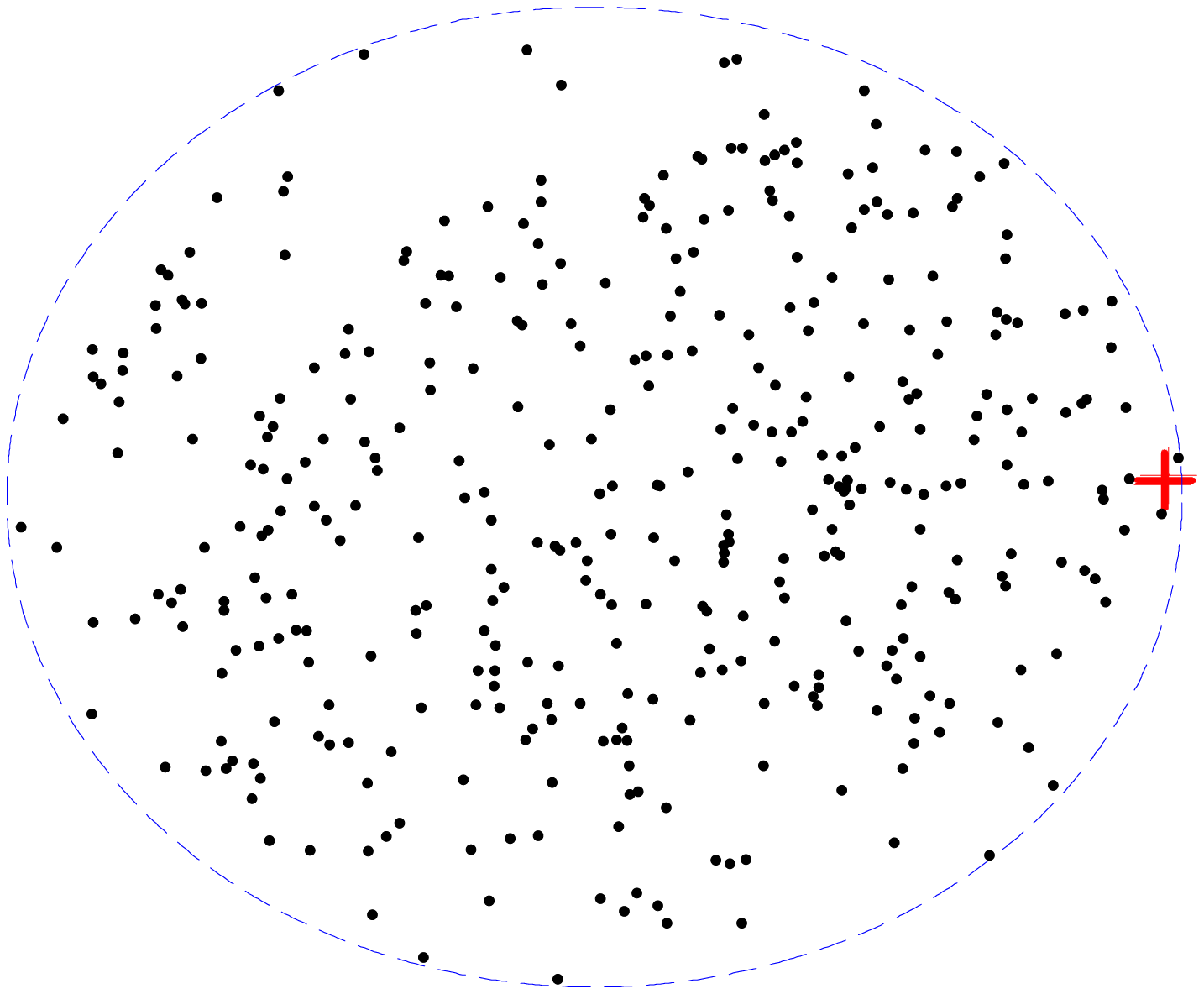}
\caption{\label{fig_z}
A 
system  
of $N=500$ , $\e=-0.6$, $\lambda=1$ viewed along (left) and perpendicular 
(right) to the axis
of rotation. The $N_c=119$ core particles are marked as $+$.}
\end{figure}

Independent of the initial conditions, all simulated systems
evolved to the same final configuration as
shown in Fig.~1.
All final (equilibrium) states 
consisted of a single core circling along the wall of the container at or 
near the equatorial plane, and a halo, also rotating. 
Systems initiated with a
single core never developed a second one, while systems initiated as rings
first fragmented into several small cores which later coalesced into bigger
ones and finally into a single core. There inevitably exists a brief period of
evolution when a ring-initiated system has two cores yet they merge
into a single core at a kinematic timescale (system crossing time) $\t=
{R^{3/2}}/{N^{1/2}}$. 
This timescale is at least by 4 orders of magnitude shorter than the time 
required for total equilibration \cite{us} and hence we cannot consider a
double-core structure even as a metastable state.

Apparently the observed single-core asymmetric
configurations are artificially excluded from consideration
in Ref.~\cite{vot} and the following publications~\cite{vot1, vot2,vot3}
by fixing the center of mass
of the system in the center of container. 
However, in a system with reflecting boundary conditions considered
in Refs~\cite{vot,vot1,vot2,vot3}, 
the total momentum is not conserved and the center of mass
does move. Once the position of the center of mass is not restricted,
high angular momentum single-core states become possible and actually
the most probable ones. Similarly to the non-rotating case,  
this can be seen by the following argument:
If two cores merge, entropy ($S$) is gained mostly due to an
increase $\D E$ in the kinetic energy of the 
high-energy halo, $\D S \approx 3/2(N-N_c)\ln(\D E)$.
Here $\D E \approx N_c^2/2r_0$ is the potential energy released
during merging of  two cores of radius $\sim r_0$  
\cite{us}
consisting of $N_c/2\sim N$ particles each. 
Entropy loss 
is negligible as  
only 3 degrees of freedom corresponding 
to the macroscopic (collective) motion of the particles of the 
vanished core become constrained.

Evidently, self-gravitating binary systems are abundant in the Universe,
while the scenario of a single core sliding along the wall is highly
unnatural. However, for a self-gravitating system to have an equilibrium
state a container is needed, and within the framework of the model with
container considered in \cite{vot}, equilibrium binary-core states do not 
exist.
 
The authors are thankful to  E.~V.~Votyakov for inspiring discussions 
and gratefully acknowledge the support
of Chilean FONDECYT under grants 1020052 and 7020052.

\end{document}